\documentclass[prl,twocolumn,superscriptaddress,showpacs]{revtex4}
\usepackage{graphicx,epsfig}
\begin{document}
\title{Time independent description of rapidly oscillating potentials}

\author{Saar Rahav}
\affiliation{Department of Physics, Technion, Haifa 32000, Israel.}
\author{Ido Gilary}
\affiliation{Department of Chemistry, Technion, Haifa 32000, Israel.}
\author{Shmuel Fishman}
\affiliation{Department of Physics, Technion, Haifa 32000, Israel.}
\date{16 July 2003}

\begin{abstract}
The classical and quantum dynamics in a high frequency field are 
found to be described
by an effective time independent Hamiltonian. It is calculated in a 
systematic expansion in the inverse of the frequency ($\omega$)
to order $\omega^{-4}$. The work is an extension of the classical
result for the Kapitza pendulum, which was calculated in the past
to order $\omega^{-2}$. The analysis makes use of an implementation
of the method of separation of time scales and of a quantum
gauge transformation in the framework of Floquet theory.
The effective time independent Hamiltonian enables one to explore
the dynamics in presence of rapidly oscillating fields,
in the framework of theories that were developed for systems
with time independent Hamiltonians.
The results are relevant, in particular, for exploration of the
dynamics of cold atoms.
\end{abstract}


\maketitle

The classical and quantum dynamics of a particle in a field that
oscillates rapidly relative to the motion of the particle
will be studied. The variation of the field in space is smooth
but otherwise arbitrary. Such fields are applied experimentally
to cold atoms, where a very high degree of control is possible.
The exploration of the dynamics of cold atoms in strong
electromagnetic fields resulted 
in many novel, interesting experimental observations~\cite{cornell02,tannudjibook2,faisal}.
The results of this letter~\cite{long}, are expected
to enable further discoveries in this field as well as in related
 fields.

For atom optics,
the effect of the internal degrees of freedom
on the center of mass motion is important.
The force on the center of mass due to the internal degrees of freedom
that couple to the external field 
is given approximately by a dipole force~\cite{tannudjibook2}. 
 The motion of the atoms can be manipulated
 by fields with amplitudes which
vary spatially, resulting in a force on the center of mass of the atoms.
In the present work, the effect of the laser on the center of mass motion is
modeled by a time dependent potential.
For some situations of physical interest
 this simpler model still
describes the dynamics of the center of mass without
the need to specify the dynamics of the internal degrees of freedom or
the quantum aspects of the light field.
In the situation that is of interest for the letter the frequency
of this potential is large compared to the inverse of the
characteristic time scale of the 
center of mass dynamics. An example of such
systems that was recently realized experimentally and that
motivated the present work is of atomic billiards~\cite{raizen01,davidson01}.
 The boundary of the billiards is 
generated by a laser beam that rapidly traverses a closed curve, which
acts as the boundary of the billiard,
approximated by the time
average of this beam. The force applied by the
boundary on the particles is approximately the mean force
applied by the beam.

The influence of a high frequency
field on a classical particle was derived by Landau
and Lifshitz~\cite{LL1} generalizing the work on the Kapitza pendulum~\cite{kapitza}, that is
 a classical pendulum with
a periodically moving point of suspension.
The motion was separated into a slow part and 
a fast part. The leading order (in the inverse frequency) of the slow motion was calculated.
This mechanism is used to trap ions in electromagnetic fields.
The most notable example is the Paul trap~\cite{paul}
that can be described approximately by a
 Hamiltonian of a time dependent harmonic oscillator~\cite{glauber95,perelomov69,perelomovbook}. 
 It is of interest to solve more general problems even if only approximately.
The work of Kapitza was first extended to quantum mechanical systems
in a pioneering paper by Grozdanov and Rakovi\'c~\cite{GR}. They introduced
a unitary gauge transformation resulting in an effective Hamiltonian
that describes the slow motion. In that paper the analysis is restricted
to a driving potential that has a particularly simple
time dependence. Moreover, the final results are restricted
to forces that are uniform in space, a situation natural
in standard spectroscopy, but too restrictive for the interesting
problems in atom optics. These restrictions are avoided in the present work.
Many investigations of periodically driven quantum systems were performed~\cite{gavrila96,vorobeichik98,henseler01,maricq82,salzman87,fernandez90,gavbook,costin}.

In the present work, a coherent theoretical treatment of
 the dynamics of classical and quantum high frequency driven
systems is presented.
Classically the motion can be considered approximately as
one that consists of a rapid oscillation around a slowly
varying trajectory. Therefore
for the classical problem the motion is separated into a ``slow'' part and a ``fast'' part and
a systematic perturbation theory is developed for the motion
of the ``slow'' part. 
 This slow motion is
computed to order $\omega^{-4}$ and demonstrated to result from an effective Hamiltonian.
It is an extension of the order $\omega^{-2}$
(that is presented in Ref.~\cite{LL1}).
Floquet theory is used to separate the slow
and fast time scales in the corresponding quantum problem.
An effective (time independent) Hamiltonian operator is defined
following \cite{GR}.
The eigenvalues of this operator are the quasienergies of 
the system. This effective Hamiltonian is then computed perturbatively
(to order $\omega^{-4}$). 
It is obtained by a gauge transformation
that is simply related to the canonical transformation leading
to the corresponding classical effective Hamiltonian.

A model Hamiltonian for the motion in a periodic field is
\begin{equation}
\label{hamilton}
H=p^2/2m + V_0(x) + V_1(x, \omega t)
\end{equation}
leading to Newton's equation  
\begin{equation}
\label{newton}
m \ddot{x} = - V_0^\prime (x) - V_1^\prime (x, \omega t),
\end{equation}
where $V_1$ is a $2 \pi$ periodic function of $\omega t$ and 
its average over a period vanishes. (We denote $\dot{X}=\frac{dX}{dt}$, $V_0^\prime=\frac{dV_0}{dx}$, etc.)
An instructive example of such a system is
\begin{equation}
\label{oscgauss}
V_0 = 0, \;\;\;\;\;\; V_1= \gamma e^{- \beta x^2} \cos (\omega t).
\end{equation}
The system is of particular interest since: 
(a) the time average of the potential
vanishes, consequently any interesting effect is due to the rapidly oscillating  potential;
(b) when $x \rightarrow \infty$
the potential vanishes and therefore one expects to find 
scattering quasienergy states.

We look for a solution of (\ref{newton}) that has the form
\begin{equation}
\label{separation}
x(t)=X(t)+\xi(X,\dot{X},\omega t)
\end{equation}
where $X$ is the slow part, $\dot{X}=\frac{dX}{dt}$ while $\xi$ is the fast part which is periodic (in the variable $\omega t$) with {\em vanishing average}.
This determines uniquely the functions $X$ and $\xi$.
One could use a different functional dependence of $\xi$ on $X$ and its
time derivatives, but such a change cannot affect the dependence
of the solution for $\xi$ on time.
Our method of solution is to choose $\xi$
so that~(\ref{newton}) leads to an equation for $X$ which is 
explicitly time independent. An exact solution using~(\ref{separation}) is too 
complicated to obtain in general. However, at high frequencies, one can determine
$\xi$ order by order in $1/\omega$, using
\begin{equation}
\label{xiexpand}
\xi = \sum_{i=1}^{\infty} \frac{1}{\omega^i} \xi_i.
\end{equation} 
The $\xi_i$ are chosen so that the equation for $X$ does not depend on $\tau \equiv \omega t$. 
One may also expand $X$ in powers of $1/\omega$ as $X=\sum_{i=0}^{\infty}  X_i/\omega^i$. When one does so 
the equation of motion for $X$ is replaced by a series of equations for $X_i$.
In this series of equations,
each $X_i$ can be determined from the lower order terms $X_j$, where $j<i$.
This is the standard method of separation
of time scales~\cite{averaging}.
These equations are equivalent, in any order, to the equation of motion of the (unexpanded) $X$ which will be used in what follows.
At a given order $\omega^{-n}$ of the present calculation {\em all} 
contributions that are found by the method of separation
of time scales are included, but {\em some} of the higher
order terms are included as well.

\begin{figure}[htb]
\includegraphics[width=7cm]{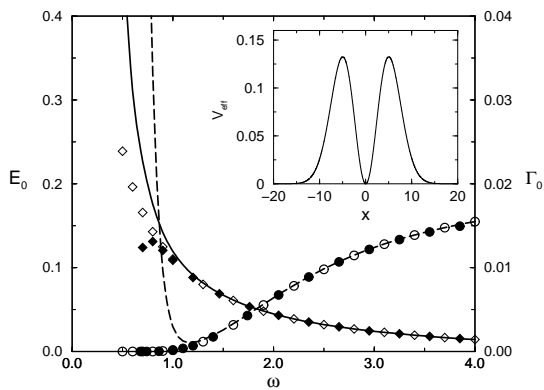}
\caption{The lowest quasienergy resonance of the oscillating Gaussian ({\protect{\ref{oscgauss}}}), $E_0$ (solid line) and $\Gamma_0$ (dashed line), as a function of the driving frequency, compared to the lowest resonance of the effective Hamiltonian ({\protect{\ref{finalg}}}), $E_0$ (diamonds) and $\Gamma_0$ (circles), for $\gamma=9$ and $\beta=0.02$ (in ``atomic units'' $\hbar=m=e=1$). Full symbols correspond to the effective Hamiltonian truncated at order $\omega^{-4}$ while empty symbols to order $\omega^{-2}$. The effective potential (\ref{veff}) is depicted in the inset, for $\omega=1.5$. \label{energy}} 
\end{figure}

The slow motion is found to be controlled by the Hamiltonian~\cite{long}
\begin{equation}
\label{classicalheff}
 H_{eff} = \frac{P^2}{2 m} + V_{eff} (X) + \frac{1}{ \omega^4} g (X) P^2 + O(\omega^{-5}),
\end{equation}
where $g (X)= \frac{3}{2m^3} \overline{\left( \int^{(2)\tau} \left[ V_1^{\prime \prime} \right] \right)^2 } $ and  $P$ is the momentum conjugate to $X$.
We define the integrals $ \int^{(j)\tau} \left[ f \right] = \underbrace{\int^\tau \left[ \cdots \int^\tau \left[ f \right] \cdots \right]}_{\mbox{$j$ times}}$, where $f(x,\tau)=\sum_{n \ne 0} f_n(x) e^{in\tau}$ and
$ \int^\tau \left[ f \right] \equiv \sum_{n \ne 0} \frac{1}{in} f_n e^{i n \tau}$.
The effective potential is found to be
\begin{equation}
\label{veff}
V_{eff} (X) \equiv V_0(X) +  V_2(X) + V_4(X).
\end{equation}
where $V_2 (X)=\frac{1}{2 m \omega^2} \overline{\left( \int^\tau \left[ V_1^\prime \right]\right)^2}$ and $V_4(X)=  \frac{1}{2 m^2 \omega^4} \left[ \overline{V_1^{\prime\prime} \left( \int^{(2)\tau} \left[ V_1^\prime \right] \right)^2} +  V_0^{\prime \prime}\overline{ \left( \int^{(2)\tau} \left[ V_1^\prime \right] \right)^2}\right]$.
For example, for the system with the potential (\ref{oscgauss}),
at the order $\omega^0$, where only the average of the potential
over time is taken into account, $V_{eff}=0$.
The effective potential $V_{eff}$ of (\ref{veff}) is plotted in
 the inset of Fig.~\ref{energy}.
The leading order ($\omega^{-2}$) is $V_2(x)=\frac{\beta^2 \gamma^2 x^2}{m \omega^2} e^{-2 \beta x^2}$, which is clearly a double barrier.
 It is obvious that it traps the particle. This
potential is always positive since it is the mean kinetic
energy of the rapid oscillation around the slow motion, an
energy which is coordinate dependent.
For this it is instructive to note that $V_2= m \overline{\dot{\xi}^2}/2$
in the leading order.
In the order $\omega^{-4}$, on the other hand, the
Hamiltonian cannot be expressed anymore in terms of an
effective potential and terms that mix coordinates and momentum,
like $g(X)P^2$ appear. Such terms result in corrections
increasing with energy. It is also more complicated to
understand terms of this order intuitively.

Consider a quantum system with a Hamiltonian that is periodic
in time, $\hat{H} (t+T)=\hat{H} (t)$. Such a system can be treated using
Floquet theory~\cite{zeldovich67,shirley65,sambe73}.
The symmetry with respect to discrete time translations
implies that the solutions of the Schr\"odinger equation
\begin{equation}
\label{schr}
i \hbar \frac{\partial}{\partial t} \psi = \hat{H} \psi
\end{equation}
are linear combinations of functions of the form
\begin{equation}
\label{quasistates}
\psi_{\lambda} = e^{-i\frac{\lambda t}{\hbar}} u_\lambda (x,\omega t)
\end{equation}
where $\lambda$ are the quasienergies and the corresponding 
quasienergy (or Floquet) states are
$u_\lambda (x, \omega (t +T))=u_\lambda (x, \omega t )$
with $\omega=2 \pi/T$.
This is the content of the Bloch-Floquet theorem in time. 
The states $u_\lambda$ are the eigenstates of the Floquet Hamiltonian
\begin{equation}
\label{floquethamiltonian}
 \hat{{\cal H}}_F = -i \hbar \frac{\partial}{\partial t} + \hat{H}.
\end{equation}

These states have a 
natural separation into a ``slow'' part $e^{-i\frac{\lambda t}{\hbar}}$
(with the natural choice $0 \le \lambda/\hbar \le \omega$), 
which includes the information about the quasienergies,
and a fast part $u_\lambda (x,\omega t)$ that depends only on
the ``fast'' time $\tau \equiv \omega t$.
In the following an equation of motion for the slow part of the
dynamics is found as was done for classical systems.
It establishes a natural link between the separation into fast and slow
motion in classical mechanics, that can be formalised by the theory
of separation of time scales, and Bloch-Floquet theory in quantum
mechanics.
For this purpose, following \cite{GR},
we will look for a unitary gauge transformation $e^{i \hat{F} (t)}$,
where $\hat{F} (t)$ is a Hermitian operator (function of $\hat{x}$ and $\hat{p}$) defined at a certain time $t$, which is a {\em periodic function} of time
with the {\em same} period as $\hat{H}$, such that in the new gauge the
Hamiltonian in the
Schr\"odinger equation is {\em time independent}.
In terms of the functions in the new gauge $\phi=e^{i \hat{F}} \psi$, the Schr\"odinger Eq. (\ref{schr}) is 
\begin{equation}
\label{schr2}
 i \hbar \frac{\partial}{\partial t} \phi= \hat{G} \phi
\end{equation}
where the Hamiltonian is
\begin{equation}
\label{effect}
\hat{G} = e^{i \hat{F}} \hat{H} e^{-i \hat{F}} + i \hbar \left( \frac{\partial e^{i\hat{F}}}{\partial t} \right)e^{-i \hat{F}}.
\end{equation}

Assume that such an operator $\hat{F}$ exists, so that $\hat{G}$ is
time independent.
Its eigenfunctions $v_\lambda (x)$ evolve as
\begin{equation}
\phi_\lambda (t,x) = e^{-i \frac{\lambda t}{\hbar}} v_\lambda (x).
\end{equation}
These states, in the original gauge, correspond to
\begin{equation}
\label{back}
\psi_\lambda (t,x)= e^{-i \hat{F}} \phi_\lambda = e^{-i \frac{\lambda t}{\hbar}}  e^{-i \hat{F}} v_\lambda (x).
\end{equation}
The function
$e^{-i \hat{F}} v_\lambda$ is periodic in time with the period of $\hat{H}$
and therefore $\psi_\lambda$ of (\ref{quasistates}) is a Floquet state with quasienergy $\lambda$
(mod $\hbar \omega$) where $u_\lambda = e^{-i \hat{F}} v_\lambda$.

At high frequencies, $\hat{F}$ is found to be small, of the order of $1/\omega$.
 We expand $\hat{G}$ and $\hat{F}$ in powers of $1/\omega$ and choose $\hat{F}$, so that $\hat{G}$ is time independent in any given order. 
The expansions are 
$ \hat{G} = \sum_{n=0}^{\infty} \hat{G}_n/\omega^n $
and
$\hat{F} = \sum_{n=1}^{\infty} \hat{F}_n/\omega^n$.
The calculation is performed by computing $\hat{G}_l$ in terms of $\hat{F}_1, \cdots, \hat{F}_{l+1}$ and then choosing $\hat{F}_{l+1}$ so that $\hat{G}_l$
is time independent.
The terms in~(\ref{effect}) are calculated with the help of the
operator expansion,
\begin{eqnarray}
\label{eifhemif}
 e^{i \hat{F}} \hat{B}  e^{-i \hat{F}} & = & \hat{B} + i \left[ \hat{F}, \hat{B} \right] - \frac{1}{2!} \left[ \hat{F},\left[ \hat{F}, \hat{B} \right] \right]
\cdots
\end{eqnarray} 
where for the first term in (\ref{effect}) one takes $\hat{B}=\hat{H}$ while for the 
second $\hat{B}=\frac{\partial}{\partial \tau}$.

The resulting time independent effective Hamiltonian is~\cite{long}
\begin{eqnarray}
\label{finalg}
\hat{G} & = & \frac{\hat{p}^2}{2m} + \hat{V}_{eff} (x) + \frac{1}{4 \omega^4} \left( \hat{p}^2 g (x) + 2 \hat{p} g (x) \hat{p} \right. \nonumber \\ & + & \left.  g (x)\hat{p}^2 \right) + \frac{\hbar^2}{\omega^4}\hat{V}_q + O(\omega^{-5}),
\end{eqnarray}
where $V_{eff}(x)$ and $g (x)$ are the classical terms (see (\ref{classicalheff}) and (\ref{veff})),
while $\hat{V}_q = \frac{1}{8 m^3} \overline{\left( \int^{(2) \tau} \left[ V_1^{(3)}\right]\right)^2}$
is a quantum correction to the classical Hamiltonian that appears first in this order. The Hamiltonian (\ref{finalg}) is presented in a form which is manifestly Hermitian.

The effective Hamiltonian~(\ref{finalg}) is the main result of this work. Its classical limit is the classical effective 
Hamiltonian (\ref{classicalheff})
that can be obtained from (\ref{hamilton}) by the canonical transformation
that is the classical limit of $-\hbar F$.
It should be emphasized that in the derivation of (\ref{finalg}) {\em no} semiclassical approximation
was made.

The perturbation theory that was developed here enables one to calculate
not only the quasienergies that are the eigenvalues of $\hat{G}$
but also the corresponding quasienergy states.
If the eigenfunctions of $\hat{G}$ are known, then the quasienergy (or Floquet) states can be 
computed up to order $\omega^{-4}$ using equation (\ref{back})
with
\begin{equation}
\label{finalf}
 \hat{F} =   \frac{1}{\hbar \omega} \int^\tau \left[ V_1 \right] + \frac{i}{m \omega^2} \int^{(2)\tau}\left( \frac{1}{2} \left[ V_1^{\prime \prime} \right]+\left[ V_1^{\prime} \right] \frac{\partial}{\partial x}\right)+ \cdots, 
\end{equation}
where the explicit expressions of terms of order $\omega^{-3}$ and $\omega^{-4}$
will be given elsewhere~\cite{long}.

The theory is demonstrated for the oscillating Gaussian (\ref{oscgauss}).
The system demonstrates trapping by an oscillating field, a phenomenon that is of
physical interest.
For this problem the effective potential (\ref{veff}),
depicted in the inset of Fig. \ref{energy}, is a double barrier
therefore it exhibits resonances. 
Each resonance is characterized by a complex energy $E-i\Gamma/2$.
(For a relevant review see~\cite{nimrev}.)
For any resonance of (\ref{finalg})  it is natural to look
for the corresponding resonance of the original time dependent Hamiltonian~(\ref{oscgauss}).
More precisely, one looks for the resonances of the Floquet 
Hamiltonian (\ref{floquethamiltonian}) with $\hat{H}$ of (\ref{oscgauss}). This is done
numerically using a combination of the ($t,t'$) method
and complex scaling~\cite{nimrev}.

The energy $E_{0}$ and the width $\Gamma_{0}$ of the lowest (smallest real part $E_0$) quasienergy-resonance
of (\ref{oscgauss}) are compared with the lowest resonance of
the corresponding effective Hamiltonian (\ref{finalg}) in Fig.~\ref{energy}.
It is clear that for large frequencies there is excellent agreement. 
The results for the effective Hamiltonian truncated
at orders $\omega^{-4}$ and $\omega^{-2}$
are comparable at large frequencies. At low frequencies the results of order $\omega^{-2}$ turn out to be more accurate,
indicating that the series do not converge at
those frequencies. For comparison the characteristic frequency
for the slow motion of the particle is in the range $0.1-0.3$.

It was found that
the perturbation theory leads to a time independent effective Hamiltonian.
This effective Hamiltonian may give physical insight, based on experience
with time independent systems, which is absent
when examining the corresponding time dependent problem.
For example, consider a system
where the time averaged potential consists of two barriers
in addition to some high frequency time dependent perturbation (with vanishing average).
If the perturbation is mainly in the region of the barriers one expects that
the perturbation slightly raises the barriers in the effective Hamiltonian.
In contrast, if the perturbation is in the region between the barriers
it tends to raise the energy of the resonance. Therefore one expects
that applying a time dependent perturbation in the region of
the barriers will tend to increase the lifetime of the resonance 
($\hbar/\Gamma$)
while applying it in between the barriers will tend to decrease it.
All the well developed techniques for time independent 
quantum systems can be used to compute the eigenvalues of $\hat{G}$,
in particular in the case where the eigenvalues and eigenstates of $\hat{G}_0=\frac{\hat{p}^2}{2m}+V_0(x)$ are known.
The effective Hamiltonian can also be used to predict trapping
by oscillating potentials that were so far investigated mainly numerically~\cite{bagwell92}.

We have investigated the dynamics of high frequency
driven general classical and quantum systems. High frequency perturbation theory, which exploits the idea of separation of time scales,
was used to obtain an effective time independent Hamiltonian for the slow part of the 
classical and quantum motion. 
The spectrum of the effective quantum Hamiltonian is the quasienergy spectrum of the
time dependent system. This effective Hamiltonian is computed to order $\omega^{-4}$ in
a perturbation theory.
While in the order $\omega^{-2}$ the effect of the rapid oscillations
around the slow motion could be expressed in terms of a classical
scalar potential, the order $\omega^{-4}$
involves both coordinates and momentum. Quantum corrections 
to the Hamiltonian also appear
at this order.

Some properties of the perturbation theory for $\hat{F}$ and $\hat{G}$,
such as its convergence and validity, are not well understood
and should be further studied. 
 For instance, in one dimension
the classical slow motion is integrable while the exact time dependent
dynamics
may have chaotic regions in phase space. This may hint that the perturbation theory
may describe correctly only part of the phase space 
(whose fraction grows with $\omega$). 

\begin{acknowledgments}
It is our great pleasure to thank M. V. Berry, N. Davidson, N. Moiseyev, and V. Rom-Kedar
for stimulating and inspiring discussions.
We thank N. Moiseyev also for the involvement in many of the fine details
of this work.
 This research was supported in part by the US-Israel Binational
Science
Foundation (BSF)
and by the Minerva Center of Nonlinear Physics of Complex Systems.
\end{acknowledgments}

\typeout{References}

\end{document}